\documentclass[11pt]{article}

\usepackage{latexsym}
\usepackage{color,graphicx}
\usepackage{amssymb}
\usepackage{amsmath}
\usepackage{amsthm}
\usepackage{txfonts}
\usepackage{enumitem}
\usepackage{fullpage}

\newtheorem{theorem}{Theorem}[section]

\newcommand{\cM}{\mathcal{M}}

\newcommand{\Mshared}[1][]{\cM \if!#1!\else (#1) \fi}

\title{On the complexity of  open shop scheduling with time lags}
\author{
Wies\l{}aw Kubiak\\
\\
\small{\emph{Faculty of Business Administration}}\\
\small{\emph{Memorial University}}\\
\small{\emph{St. John's, Canada}}
}
                                    
\begin{document}
\maketitle
\begin{abstract}
The minimization of makespan in open shop with time lags has been shown NP-hard in the strong sense even for the case of two machines and unit-time operations. The minimization of total completion time  however has remained open for that case though 
it has been shown NP-hard in the strong sense for weighted total completion time or for jobs with release dates. This note shows that the minimization of total completion time for two machines and unit-time operations is NP-hard in the strong sense which answers the long standing open question.
\end{abstract}


\section{Introduction}

The open shop with job-dependent time lags has been studied for quite sometime in the literature. The time lags model  delays required between job's operations  due to necessary transportation needed to move a job from one machine to another for instance. Zhang \cite{Z10} provides
an interesting discussion of the time lag models and their applications in scheduling. Most research on the open shops with time lags has focused on two machine open shops where each job $J_i$, $i=1, \dots, n$, consists of two operations $O_{i1}$ and $O_{i2}$ to be processed on two machines $M_1$ and $M_2$ respectively in any order. The operations $O_{i1}$  and $O_{i2}$ processing times  equal $p_{i1}> 0$ and $p_{i2}> 0$ respectively, and the time lag is $l_i\geq 0$. In a feasible schedule either machine can process at most one job at a time, each job can be processed by at most one machine at a time, and the \emph{later} operation of job $J_i$  in the schedule needs to wait at least $l_i$ time units to start  following the completion of the \emph{earlier} operation of job $J_i$ in the schedule. Yu \cite{Y96}, and Yu et al. \cite{YHL04} prove
a series of strong complexity results for  makespan minimization. They prove that the problem is NP-hard in the strong sense even if all operations are unit-time operations. This problem is denoted by $O2|p_{ij}=1, l_{i}|C_{\max}$ in the well-known notation of Graham et al \cite{GLLRK79}.
Yu \cite{Y96} then goes on to prove that the problem is NP-hard in the strong sense even if there are only two possible values $l$ and $l'$ of time lags in a job-proportionate open shop, i.e. the problem $O2| p_{i1}=p_{i2}, l_i\in \{l,l'\}| C_{max}$, and it is also NP-hard in the ordinary sense when only one value $l$ of time lag is permitted in a job-proportionate open shop, i.e. the problem $O2| p_{i1}=p_{i2}, l_i=l| C_{max}$. Rebain and Strusevich   \cite{RS99}  give a linear time algorithm for the instances with short time lags, i.e time lags that meet the following
condition $\max_i\{l_i\}\leq \min_{ij}\{p_{ij}\}$. These results determine current borderline between NP-hard and polynomially solvable cases for the two machine open shop makespan minimization problem with job dependent-time lags. The problem intractability caused
research to focus on approximation algorithms and on-line competitive algorithms for makespan minimization. Strusevich  \cite{S99} gave $\frac{3}{2}$- approximation algorithm, and Zhang and van de Velde  \cite{ZV10} gave a 2-competitive algorithm for $O2|l_i| C_{max}$.
The reader is referred to Zhang \cite{Z10}  for a comprehensive review of approximation and on-line algorithms for the problem

Brucker et al. \cite{BKCS04} switch attention to other than makespan objective functions. In particular to the total completion time objective which is another key scheduling objective function. They prove that \emph{weighted} total completion time minimization is NP-hard in the strong sense, i.e. the problem $O2| p_{ij}=1, l_i|\sum w_iC_i$. They prove that the same holds for the total completion time with jobs being \emph{released} possibly at different times, i.e. the problem $O2| p_{ij}=1, l_i, r_i| \sum C_i$. In this paper we prove that  the problem where all jobs are released at the same time and their weights are all equal, i.e.
the problem $O2| p_{ij}=1, l_i| \sum C_i$ is NP-hard in the strong sense. This result strengthens those earlier complexity results for total completion time, and it answers a question that has been open at least since the paper by Brucker et al. \cite{BKCS04}. The prove is given in the next section.


\section{NP-hardness proof}

Let $l_1, \dots, l_n, e$ be non-negative integers, and $n$ a positive \emph{even} integer such that $n<e< \frac{3n}{2}$.  Let  $A$ and $B$ be a partition of the index set $\{1, \dots, n\}$ into two disjoint sets of equal size $\frac{n}{2}$. 
For simplicity denote $\{\ell_1, \dots, \ell_{\frac{n}{2}}\}=\{l_j : j\in A\}$ and  $\{\lambda_1, \dots, \lambda_{\frac{n}{2}}\}=\{l_j : j\in B\}$. Consider the following question.

(Q) Is there a partition of the set $\{1, \dots, n\}$ into two disjoint sets $A$ and $B$ of equal size $\frac{n}{2}$ such that there is a permutation $\pi_A$ of the set $\{1,\dots, \frac{n}{2}\}$
and a permutation $\sigma_A$ of the set $\{\frac{n}{2}+1, \dots, n\}$ satisfying
\begin{equation}
\pi_A(i)+\ell_i+\sigma_A(i)=e
\end{equation}
for $i=1,\dots, \frac{n}{2}$, and
there is a permutation $\pi_B$ of the set $\{\frac{n}{2}+1, \dots, n\}$
and a permutation $\sigma_B$ of the set $\{1, \dots, \frac{n}{2}\}$ satisfying
\begin{equation}
\pi_B(i)+\lambda_i+\pi_A(i)=e
\end{equation}
for $i=1,\dots, \frac{n}{2}$? 

We refer to this problem as Partition Restricted Numerical 3-Dimensional Matching (PRN3DM) problem. It is easy to verify that any instance of the PRN3DM problem with an affirmative answer to Q must satisfy the following condition

\begin{equation}\label{E1}
\Sigma_{i=1}^n l_i=n(e-n-1),
\end{equation}
therefore without loss of generality we limit the  PRN3DM to the instances for which (\ref{E1}) holds. The problem PRN3DM is NP-hard in the strong sense which follows from Theorem 1 on p. 30 in Yu \cite{Y96}.

The corresponding instance of the decision open shop problem $O2|p_{ij}=1, l_i|\Sigma C_i\leq F$ is made up of $n$ jobs $J_1, \dots, J_n$ with time lags $L_1=l_1+\Delta, \dots, L_n=l_n+\Delta$ respectively, where $\Delta=\frac{3n}{2}-e$. The threshold for total completion time 
equals $F=\frac{n}{2}(\frac{3n}{2}+1)$. 

\begin{figure}
\centering         
\includegraphics[scale=0.60]{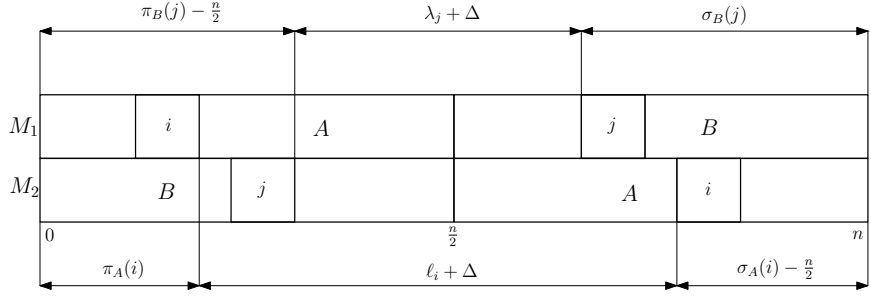}
\caption{Schedule S for the partition $A$ and $B$ and permutations $\pi_A$, $\sigma_A$, $\pi_B$, and $\sigma_B$.}
\label{openshopA}
\end{figure}

For a partition $A$ and $B$ and permutations $\pi_A$, $\sigma_A$, $\pi_B$, and $\sigma_B$ that attest to an affirmative answer to Q, the schedule $S$ in Figure \ref{openshopA} is a feasible schedule for the open shop problem with total completion time 
equal to

\begin{equation*}
\Sigma_{i=1}^{\frac{n}{2}} (\pi_A(i)+\ell_i +\Delta+1) + \Sigma_{i=1}^{\frac{n}{2}} (\pi_B(i)-\frac{n}{2}+\lambda_i +\Delta+1),
\end{equation*}
which by (\ref{E1}) and definition of $\Delta$ equals $F$. Thus $S$ gives an affirmative answer to the problem $O2|p_{ij}=1, l_i|\Sigma C_i\leq F$ instance.


Now, let $\mathcal{S}$ be a feasible schedule for the instance of $O2|p_{ij}=1, l_i|\Sigma C_i\leq F$ with total completion time not exceeding $F$. We first show that the makespan $C_{\max}=n$ in $\mathcal{S}$. To that end let $x_{\sigma(1)}\leq x_{\sigma(2)}\leq \dots \leq x_{\sigma(n-1)} \leq x_{\sigma(n)}$ be the times when the \emph{earlier} operations of the jobs
$J_1, \dots, J_n$ complete in $\mathcal{S}$.  Because of the delay due to the time lags the total completion time of $\mathcal {S}$  is at least
\begin{equation*}
\sum_{i=1}^{\frac{n}{2}} (x_{\sigma(2i-1)}+x_{\sigma(2i)}) + \sum_{j=1}^n (L_j +1),
\end{equation*}
which does not exceed the threshold $F$ for $\mathcal{S}$. Hence by (\ref{E1}) and definition of $\Delta$
\begin{equation} \label{Ch11start}
\sum_{i=1}^{\frac{n}{2}} (x_{\sigma(2i-1)}+x_{\sigma(2i)})\leq \frac{n}{2}(\frac{n}{2}+1).
\end{equation}
For two machines we have $i\leq x_{\sigma(2i-1)} \leq x_{\sigma(2i)}$, $i=1, \dots, \frac{n}{2}$. Thus by (\ref{Ch11start}) we get $x_{\sigma(2i-1)}=x_{\sigma(2i)}=i$ for $i=1, \dots, \frac{n}{2}$.  Therefore each job $J_1, \dots, J_n$ completes after time $\frac{n}{2}$ in 
$\mathcal{S}$. Let $C_{\pi(1)}\leq C_{\pi(2 )}\leq \dots \leq C_{\pi(n-1)} \leq C_{\pi(n)}$ be the completion times of the  jobs
$J_1, \dots, J_n$  in $\mathcal{S}$.  Clearly $C_{\pi(i)}=\frac{n }{2}+ c_{\pi(i)}$, for some $c_{\pi(i)} \geq 1$, thus
\begin{equation} \label{Ch11complete}
\sum_{i=1}^{\frac{n}{2}} (c_{\pi(2i-1)} + c_{\pi(2i)}) \leq \frac{n}{2}(\frac{n}{2}+1)
\end{equation}
in $\mathcal{S}$. Again, for two machines we have $i\leq c_{\pi(2i-1)} \leq c_{\pi(2i)}$, $i=1, \dots, n$. Thus by (\ref{Ch11complete}) we get $c_{\pi(2i-1)}=c_{\pi(2i)}=i$ for $i=1, \dots, \frac{n}{2}$. Therefore all jobs complete by $C_{\max}=n$ in $\mathcal{S}$ which is what we set out to show first.
\begin{figure}
\centering         
\includegraphics[scale=0.60]{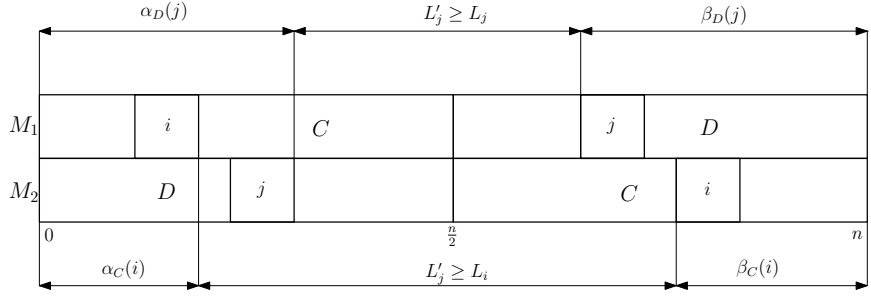}
\caption{Schedule S with total completion time not exceeding F.}
\label{fig3}
\end{figure}

Finally, let $C$ be the set of $\frac{n}{2}$ jobs with earlier operations in the interval $[0,\frac{n}{2}]$ on $M_1$ and later operations in $[\frac{n}{2},n]$ on $M_2$ in $\mathcal{S}$, and $D$ be the set of $\frac{n}{2}$ jobs with earlier operations in the interval $[0,\frac{n}{2}]$ on $M_2$ and later operations in $[\frac{n}{2},n]$ on $M_1$ in $\mathcal{S}$, see Figure \ref{fig3}. We have
\begin{equation*}
\alpha_C(i) +L'_i+\beta_C(i)=n
\end{equation*}
for each $i\in C$, and
\begin{equation*}
\alpha_D(j) +L'_j+\beta_D(j)=n
\end{equation*}
for each $j\in D$ for some permutations $\alpha_C$, $\alpha_D$, $\beta_C$, $\beta_D$ of the set $\{1,\dots, \frac{n}{2}\}$. Thus
\begin{equation*}
\alpha_C(i) +L'_i+\beta_C(i)+\frac{n}{2}=\frac{3n}{2}
\end{equation*}
for each $i\in C$, and
\begin{equation*}
\alpha_D(j) +\frac{n}{2}+L'_j+\beta_D(j)=\frac{3n}{2}
\end{equation*}
for each $j\in D$. By taking  the permutation $\pi_C=\alpha_C$ of $\{1, \dots, \frac{n}{2}\}$ and $\sigma_C=\beta_C+ \frac{n}{2}$ of  $\{ \frac{n}{2}+1, \dots, n\}$,
and  the permutation $\pi_D=\alpha_D+ \frac{n}{2}$ of $\{ \frac{n}{2}+1, \dots, n\}$ and $\sigma_D=\beta_D$ of  $\{1, \dots, \frac{n}{2}\}$, we get
\begin{equation}\label{E2}
\pi_C(i) +l'_i+\sigma_C(i)=\frac{3n}{2}-\Delta=e
\end{equation}
for each $i\in C$, and
\begin{equation}\label{E3}
\pi_D(j) +l'_j+\sigma_D(j)=\frac{3n}{2} -\Delta=e
\end{equation}
for each $j\in D$, where $L'_i=l'_i+\Delta$ and $l'_i\geq l_i$ for $i=1, \dots, n$. By (\ref{E2}) and (\ref{E3}) we have
\begin{equation*}\label{E0}
\Sigma_{i=1}^n l'_i=n(e-n-1).
\end{equation*}
Thus by (\ref{E1}), $l'_i=l_i$ for $i=1, \dots, n$. Hence
\begin{equation}
\pi_C(i) +l_i+\sigma_C(i)= e
\end{equation}
for each $i\in C$, and
\begin{equation}
\pi_D(j) +l_j+\sigma_D(j)= e
\end{equation}
for each $j\in D$. Therefor $C$ and $D$ make up the required partition, which proves the following theorem.

\begin{theorem}
The problem $O2|p_{ij}=1, l_i|\sum C_i$ is NP-hard in the strong sense.
\end{theorem}

\bibliographystyle{plain}
\bibliography{BibChapter1a}

\end{document}